\documentstyle[aps,preprint]{revtex}
\begin{document}
\draft
\preprint{SOGANG-HEP-216/96}
\title{Entropy in the Kerr-Newman Black Hole}
\author{
Jeongwon Ho\footnote{E-mail address : jwho@physics.sogang.ac.kr},
Won T. Kim\footnote{E-mail address : wtkim@ccs.sogang.ac.kr}, and 
Young-Jai Park\footnote{E-mail address : yjpark@ccs.sogang.ac.kr}
}
\address{
Department of Physics and Basic Science Research Institute,\\
Sogang University, C.P.O.Box 1142, Seoul 100-611, Korea\\
}
\author{Hyeonjoon Shin\footnote{E-mail address : hshin@ctp.snu.ac.kr}}
\address{
Department of Physics, Seoul National University, Seoul 151-742, Korea}
\date{March 1997}
\maketitle
\begin{abstract}
Entropy of the Kerr-Newman black hole is calculated via the 
brick wall method with maintaining careful attention to the contribution
of superradiant scalar modes.  It turns out that the nonsuperradinat
and superradiant modes simultaneously contribute to the 
entropy with the same order in terms of the brick wall cutoff $\epsilon$. In particular, the contribution of the superradiant modes to the entropy is negative.  It is shown that $\theta$-dependence structure of the brick wall 
cutoff is given by a lower bound of the angular velocity and naturally requires an angular cutoff $\delta$.  Finally, if the cutoff values, $\epsilon$ and $\delta$, satisfy a proper relation between them, the resulting entropy satisfies the area law.
\end{abstract}
\bigskip
 
\newpage
  
\section{INTRODUCTION}
Since Bekenstein and Hawking's pioneering works[1,2] that the entropy of a black hole is proportional to its surface area were presented, many 
efforts have been devoted to study the statistical origin of the black 
hole entropy[3-8].  One of the endeavors is the {\it brick wall model}[8] by 
't Hooft in which the black hole entropy is identified with the entropy of 
a thermal gas of quantum field excitations outside the event horizon. In the model, a brick wall cutoff (a fixed boundary at near event 
horizon) is introduced in order to eliminate divergences arising from
the infinite 
growth of the density of states close to the horizon.  This method has been
applied to the various black holes[9-15]. 

The application for the brick wall method to the rotating black hole is not simple comparing with the non-rotating cases[13, 14]. The difficulties are originated from the fact that the scalar field on the rotating 
black hole has two kinds of modes : 
superradiant(SR) and nonsuperradiant(NSR) 
modes[16].  Then the free energy is composed of the SR part $F_{SR}$
and the NSR one $F_{NSR}$.  In general, $F_{SR}$ is divergent due to large 
azimuthal quantum number.  According to this reason, the authors of ref.[13] in which the Kerr-Newman (KN) black hole[17] entropy is calculated by means of the brick wall method,
only considered the NSR part. In ref.[14], however, it has been shown that 
in the (2+1) dimensional rotating black hole,
the divergence of $F_{SR}$ can be regularized by introducing a suitable
cutoff on the azimuthal quantum number $\epsilon_m$ and the 
whole entropy is given by NSR and SR 
contributions.  Fortunately, the additional 
cutoff $\epsilon_m$ does not 
appear in the entropy. Moreover, the fact that there is a contribution of SR modes gives us some new features of the brick wall method as follows ;  The non-rotating limit in the brick wall method for the rotating black holes is meaningless, and the non-r
otating and the rotating cases should be treated separately. In this situation, one is to obtain a lower bound of the angular velocity from a condition to preserve the validity of the brick wall method for the rotating black hole.

In this paper, we shall reconsider the application for the KN black hole to 
the brick wall method.  There are some different properties in the KN black hole in contrast with the (2+1) rotating black hole. In the (2+1) dimensional rotating black hole case, the SR modes give the sub-leading order contribution to the entropy in term
s of the brick wall cutoff $\epsilon$ while the NSR modes give the leading order one, and thus SR effect is minor. In the KN black hole case, however, the leading order contribution to the entropy is given by the SR parts as well as the NSR ones. The cont
ribution of SR modes is negative. On the other hand, the brick wall cutoff in the KN black hole background is dependent on $\theta$. It is shown that the $\theta$-dependence structure of the brick wall cutoff can be obtained from the lower bound of the an
gular velocity and naturally requires the angular cutoff $\delta$.  Without loss of generality, we could choose the brick wall cutoff $\epsilon$ and the angular cutoff $\delta$ such that
the entropy obtained in the brick wall method should satisfy the area law.

  In sect. II, the free energy of the scalar field on the KN black
hole is calculated in terms of the brick wall method. And the non-rotating
limit and the $\theta$-dependence structure of the brick wall
cutoff are discussed.
The entropy of the scalar field is obtained in sect. III. In sect. IV, the contents of this paper is briefly summarized. 

\section{FREE ENERGY}  
In this section, the free energy of the scalar field around the KN black 
hole is calculated in terms of the brick wall method.
     The line element of the background spacetime is given by
\begin{eqnarray}
 ds^{2}=&g_{tt}&dt^{2} + 2g_{t\phi}dtd\phi + g_{\phi\phi}d\phi^{2}
          + g_{rr}dr^{2} + g_{\theta\theta}d\theta^{2} \nonumber \\
       =
  &-&\frac{\Delta - a^{2}\sin^{2}\theta}{\Sigma}dt^{2} 
        - \frac{2a\sin^{2}\theta(r^{2} +a^{2} -\Delta)}{\Sigma}dtd\phi
                    \nonumber \\
  &+& \left[\frac{(r^{2} +a^{2})^{2} - \Delta a^{2}\sin^{2}\theta}{\Sigma}
       \right]
      \sin^{2}\theta d\phi^{2}   
  +\frac{\Sigma}{\Delta}dr^{2} + \Sigma d\theta^{2},
\end{eqnarray}
where $\Sigma (r,\theta) \equiv r^{2} + a^{2}\cos^{2}\theta$, $\Delta (r) 
\equiv r^{2} -2Mr +a^{2} + e^{2}$, and $M$, $a$, and $e$ are the mass, 
the angular momentum per unit mass, and the charge of the black hole, 
respectively.  The KN black hole has two coordinate singularities 
corresponding to the outer and inner horizon 
$r_{\pm}=M \pm \sqrt{M^{2} - a^{2} - e^{2}}$, where 
$M^{2} \geq a^{2} + e^{2}$
and the equality holds in the extremal case, $r_{+} = r_{-}$.
The event horizon is defined by $r_{H}=r_{+}$.  The KN metric (1) has  
a stationary limit surface at $r_{erg}=M + \sqrt{M^{2} - a^{2}\cos^{2}\theta 
-e^{2}}$ called ergosphere in which a particle can not remain at rest as 
viewed from infinity.   

  Let us now begin with the Klein-Gordon equation of the scalar field 
on the KN black hole background
\begin{eqnarray}
\frac{1}{\sqrt{-g}}
\partial_{\mu}(\sqrt{-g}\partial^{\mu}\Phi) -\mu^{2}\Phi =0,
\end{eqnarray}
where $\mu$ is the mass of a scalar field $\Phi$.  The WKB approximation with
$\Phi (t, r, \theta , \phi )=\exp[-iEt +im\phi + iK(r, \theta)]$ to the 
Klein-Gordon equation yields the following constraint[8, 10] 
\begin{eqnarray}
k_{r}^{2} = \frac{\Sigma}{(\Delta \sin\theta)^{2}}[g_{ab}T^{a}T^{b}
      - \Delta(\Sigma^{-1}k_{\theta}^{2} + \mu^{2})\sin^{2}\theta ],
\end{eqnarray}
where $T^{t} \equiv m$, $T^{\phi} \equiv E$, $g_{ab}T^{a}T^{b} = 
g_{\phi\phi}E^{2} + 2g_{t\phi}Em +g_{tt}m^{2}$
and indices $a,~b$ denote $t$ and $\phi$ coordinates.
$m$ is an azimuthal quantum number and $k_{\mu} \equiv \partial_{\mu}K$.  
Then, according to the semiclassical quantization rule, the radial wave 
number is given by
\begin{eqnarray}
\pi n_{r} =
       \int d\theta d\phi \int^{L}_{r_{H}+\epsilon}dr \int dk_{\theta}k_{r},
\end{eqnarray} 
where $\epsilon$ is a small, positive quantity representing an ultraviolet
cutoff and $L$ is an infrared cutoff.  Note that since the background 
spacetime, 
KN black hole, has the axial symmetry,
 the cutoff $\epsilon$ could be dependent on $\theta$. 
In eq.(4), the integration 
with respect to  $k_{\theta}$ must be carried out over the phase volume 
satisfying the following condition 
\begin{eqnarray}
 g_{ab}T^{a}T^{b} - \Delta \mu^{2} \sin^{2}\theta \geq 0,
\end{eqnarray}
due to $k_{r}^{2} \geq 0 $, and the radial wave number becomes
\begin{eqnarray}
\pi n_{r} =
        \frac{\pi}{2}\int d\theta d\phi \int^{L}_{r_{H}+\epsilon}dr 
        \frac{\Sigma}{\Delta^{3/2} \sin^{2}\theta }[g_{ab}T^{a}T^{b}
            -  \Delta\mu^{2} \sin^{2}\theta ].
\end{eqnarray}  
Then the free energy $F$ at the inverse temperature $\beta$ 
on the rotating black hole is given by[14, 16]
\begin{eqnarray}
\beta F &=& {\displaystyle \sum_{K}} \ln\left[1-e^{-\beta(E_{K}-m\Omega_{H})}
                      \right]  
     \approx  \int dn_{r} \int dm \ln\left[1-e^{-\beta(E-m\Omega_{H})}
                      \right]  \nonumber \\
      &=& - \beta \int dm \int dE \frac{n_{r}}{
             e^{\beta(E-m\Omega_{H})} - 1}  \nonumber  \\
    &=& -\frac{\beta}{2}\int d\theta d\phi \int_{r_{H}+\epsilon}^{L}dr 
    \frac{\Sigma}{\Delta^{3/2} \sin^{2}\theta }\int dm \int dE
     \frac{g_{ab}T^{a}T^{b} - \Delta\mu^{2} \sin^{2}\theta }{
        e^{\beta(E-m\Omega_{H})} - 1},
\end{eqnarray}  
where the approximation denotes the continuum limit and we have integrated
by parts in the second line in eq.(7).  The angular velocity of the black
hole horizon is defined by
\begin{eqnarray}
\Omega_{H} = -\displaystyle \left.\frac{g_{t\phi}}{g_{\phi\phi}
              } \right|_{r=r_{H}}  = \frac{a}{r_{H}^{2}+a^{2}}~~. 
\end{eqnarray} 
In eq.(7), we assumed that the scalar field is rotating with the angular
velocity $\Omega_H$, i.e., is in the Hartle-Hawking vacuum state[18].

    As above mentioned, the free energy can be separated into two parts as 
$F=F_{NSR} + F_{SR}$.  The SR modes are characterized by the condition of 
$0 \leq E \leq m \Omega_{H}$ and $m > 0$, while the NSR modes are 
obtained with $E > m \Omega_{H}$ for arbitrary $m$[16].
In these integration ranges, we now calculate the free energy of the NSR
and the SR modes.

\subsection{Free Energies of NSR and SR Modes}
In eq.(7), the NSR part of the free energy $F_{NSR}$ is given by
\begin{eqnarray}
F_{NSR} = -\frac{1}{2}\int d\theta d\phi \int_{r_{H}+\epsilon}^{L}dr 
 \frac{\Sigma}{\Delta^{3/2} \sin^{2}\theta }\int dm \int_{m\Omega_{H}}^{
\infty}
 dE\frac{g_{ab}T^{a}T^{b} - \Delta\mu^{2} \sin^{2}\theta }{
        e^{\beta(E-m\Omega_{H})} - 1}~.
\end{eqnarray}   
Under the condition for the finiteness of the free energy and eq.(5),
the free energy (9) becomes as follows[13]
\begin{eqnarray}
F_{NSR} \approx  - \frac{\zeta(4)}{\beta^{4}} \int d\theta d\phi 
\left[  
\frac{(r_{H}^{2} + a^{2})^{4}\sin\theta}{(r_{H}^{2} + a^{2}\cos^{2} \theta)
 (r_{H} - M)^{2}} \frac{1}{\epsilon} +
  f_{1}(\theta)\ln\left(\frac{r_{H}}{\epsilon}\right) \right]
 + {\cal O}(\epsilon),
\end{eqnarray}
where
\begin{eqnarray}
f_{1}(\theta) &\equiv& -\frac{(r_{H}^{2}+a^{2})^{4}\sin\theta }{
 (r_{H}^{2} +a^{2}\cos^{2}\theta)^{3}(r_{H}-M)^{3}}         \nonumber \\
              & &  \times \left[  (r_{H}^{2} +a^{2}\cos^{2}\theta)^{2} 
                 -6r_{H}(r_{H}-M) (r_{H}^{2} +a^{2}\cos^{2}\theta)
                 - 4a^{2}r_{H}^{2}\sin^{2}\theta
                         \right].
\end{eqnarray}
The approximation in eq.(10) means that we used the small mass limit and
made the integration 
w.r.t. the radial coordinate $r$ near the horizon.

Let us now consider the remaining SR part of the free energy
$F_{SR}$ which is given by
\begin{eqnarray}
F_{SR} &=& -\frac{1}{2}\int d\theta d\phi \int_{r_{H}+\epsilon}^{L}dr 
 \frac{\Sigma}{ \Delta^{3/2}\sin^{2}\theta}\int_{>0} dm \int^{m\Omega_{H}
}_{0}
 dE\frac{g_{ab}T^{a}T^{b} - \mu^{2}\Delta \sin^{2}\theta}{
e^{\beta(E-m\Omega_{H})} - 1}
             \nonumber  \\
&\approx & -\frac{\Omega_{H}}{2}\int 
d\theta d\phi \int_{r_{H}+\epsilon}^{L}dr 
 \frac{\Sigma}{\Delta^{3/2}\sin^{2}\theta}\int_{>0} dm m^{3} \int_{0}^{1}dx
  \frac{ g_{ab}\Omega^{a} (x)\Omega^{b} (x)}{
       e^{-\beta m \Omega_{H}(1-x)} - 1},
\end{eqnarray}
where we performed the change of variable, $E \rightarrow xm\Omega_{H}$, and
\begin{eqnarray}
&& g_{ab}T^{a}T^{b} \rightarrow m^2g_{ab}\Omega^{a} (x)
  \Omega^{b} (x) \equiv m^2\Omega^{2}(r, \theta ;x)
  \equiv m^2(g_{\phi\phi}\Omega_{H}^{2} x^{2} + 2g_{t \phi}\Omega_{H} x + g_{tt}),
\nonumber \\
&& \Omega^{t} \equiv 1,~~\Omega^{\phi} \equiv \Omega_{H}x .
\end{eqnarray}
The approximation in eq.(12) denotes that we take the small mass limit. 

Like the (2+1) dimensional rotating black hole case[14], due to the $m$-integration in eq.(12), the SR part of the free energy becomes divergent. Following the prescription of ref.[14], as we introduce a regulating factor $e^{-\epsilon_{m} m}$, the $m$-in
tegration is regulated
as follows  
\begin{eqnarray}
F_{SR} &=& -\frac{\Omega_{H}}{2}\int d\theta d\phi 
\int_{r_{H}+\epsilon}^{L}dr 
\frac{\Sigma}{\Delta^{3/2}\sin^{2}\theta}
 \int_{0}^{1}dx \int_{>0} dm m^{3} 
  \frac{\Omega^{2} (r, \theta ;x)e^{-\epsilon_{m} m}}{
       e^{-\beta m \Omega_{H}(1-x)} - 1}
             \nonumber  \\
 &=& \frac{\Omega_{H}}{2}\int d\theta d\phi \int_{r_{H}+\epsilon}^{L}dr 
 \frac{\Sigma}{\Delta^{3/2}\sin^{2}\theta} \int_{0}^{1}dx
 \Omega^{2}(r, \theta ;x)
\left[
  \frac{6}{\epsilon_{m}^{4}} + \frac{6\zeta (4)}{[\beta \Omega_{H}(1-x)]^{4}}
          \right].
\end{eqnarray}
Note that $F_{SR}$ (14) is obviously divergent as the regulating infinitesimal parameter $\epsilon_{m}$ goes to zero. 

Now, consider the $x$-integration in eq.(14). The integrand $(1-x)^{-4}$ term entered in the $x$-integration diverges at $x=1$. However, in order to remove divergence, we does not need to introduce another regulating factor. Instead, it can be easily reme
died by considering the condition (5) as follows ; Upon the small mass limit and the considering change of variable,
the condition (5) is rewritten by
\begin{eqnarray}
g_{\phi\phi}\Omega_{H}^{2} x^{2} 
+ 2g_{t \phi}\Omega_{H} x + g_{tt} \geq 0.
\end{eqnarray}
In the region $r> r_{erg}$, metric components become $ g_{\phi \phi}>0,~~ g_{t \phi}<0$ and $g_{tt}<0$, then the condition (15) is not satisfied in the range $0 <x<1$.  In the region 
$r_{H} <r< r_{erg}$, since metric components become $ g_{\phi \phi}>0,~~ g_{t \phi}<0$ and $g_{tt}>0$, the condition (15) is valid only in the range $0< x < \alpha$, where $\alpha \equiv - (g_{t\phi}+\Delta^{1/2}\sin\theta)/
(g_{\phi \phi}\Omega_{H})$ and $\alpha < 1$. According to this restriction, we must carry out the $x$-integration from $0$ to $\alpha$ in the space range 
$r_{H} <r< r_{erg}$. Thus the  free energy of SR modes (14) becomes
\begin{equation}
F_{SR} = \frac{F_1}{\epsilon^{4}_{m}} + F_2,
\nonumber
\end{equation}
where
\begin{eqnarray}
F_1 &\equiv& \int d\theta d\phi 
	\int_{r_{H}+\epsilon}^{r_{erg}}dr 
	\frac{\Sigma}{\Delta^{3/2}\sin^{2}\theta}
        \left[
	     \frac{g_{\phi\phi}}{3}(\Omega_{H}\alpha)^{3} 
              + g_{t \phi}(\Omega_{H}\alpha)^{2} + g_{tt}\Omega_{H}\alpha 
        \right]        
\nonumber  \\
F_2 &\equiv& \frac{\zeta (4)}{\beta^{4}\Omega_{H}^{3}}\int d\theta d\phi 
  \int_{r_{H}+\epsilon}^{r_{erg}}dr 
 \frac{\Sigma}{\Delta^{3/2}\sin^{2}\theta} 
  \left[
    (-g_{\phi\phi}\Omega_{H}^{2} + g_{t \phi}\Omega_{H}  - g_{tt}) 
     + \frac{3g_{\phi\phi}\Omega_{H}^{2}}{1- \alpha}
  \right.       
\nonumber    \\ 
 && \left. 
   - \frac{3(g_{\phi\phi}\Omega_{H}^{2} + g_{t \phi}\Omega_{H})}{( 1- \alpha)^2}
   +\frac{(g_{\phi\phi}\Omega_{H}^{2}
          +2g_{t \phi}\Omega_{H} + g_{tt})}{( 1- \alpha)^3} 
    \right].
\end{eqnarray}
The fact that the upper bound of the $r$-integration range is restricted by
$r_{reg}$ makes us understand an important characteristic of the brick wall method for the rotating black holes.  We shall minutely discuss about that in subsection B.

Finally, performing the $r$-integration near horizon in eq.(17), $F_{SR}$ is approximately given by
\begin{eqnarray}
F_{SR} &\approx & \frac{\zeta(4)}{\beta^{4}} \int d\theta d\phi
  \left[
  \frac{(r_{H}^{2} + a^{2})^{4}\sin\theta }{
2(r_{H}^{2} + a^{2}\cos^{2} \theta)
 (r_{H} -M)^{2}} \frac{1}{\epsilon} 
  + \frac{f_{2}(\theta)}{\sqrt{\epsilon}} 
  + \frac{f_{1}(\theta)}{2}\ln\left(\frac{r_{H}}{\epsilon}\right) \right]
    + \frac{F_{1}}{\epsilon^{4}_{m}}
        \nonumber  \\
 &\approx & - \frac{1}{2} F_{NSR} + \frac{\zeta(4)}{\beta^{4}} 
    \int d\theta d\phi \frac{f_{2}(\theta)}{\sqrt{\epsilon}} 
    + \frac{F_{1}}{\epsilon^{4}_{m}},
\end{eqnarray}
where
\begin{eqnarray}
f_{2}(\theta) &\equiv &
-\frac{3\sqrt{2}(r_{H}^{2} +a^{2})^{4}}{
     2a(r_{H}-M)^{5/2}(r_{H}^{2}+a^{2}\cos^{2}\theta)^{2}}
  \left[
 (r_{H} -M)(r_{H}^{2}+a^{2}\cos^{2}\theta) + r_{H}a^{2}\sin^{2}\theta 
    \right].
\end{eqnarray}
   
Let us now compare the divergence structures of $F_{NSR}$ and $F_{SR}$ in
terms of the brick wall cutoff $\epsilon$.  In eqs.(10) and (18), the leading order of $F_{SR}$ is equal to that of $F_{NSR}$ as $\epsilon^{-1}$. Note that in (2+1) dimensional rotating black hole, the contribution of $F_{SR}$ to the total free energy is 
sub-leading order[14]. Interestingly, the linear and logarithmic divergence terms of $F_{SR}$
is minus a half times those of $F_{NSR}$ in eq.(18).  In other words, the SR modes negatively contribute to the total free energy. We shall discuss the physical meaning of the negative contribution
in Section III.

\subsection{Non-rotating limit and $\theta$-dependence structure of $\epsilon$}   
Since we now have the expression of the free energy, which shows the
divergence structure in term of the brick wall cutoff explicitly, we may proceed to calculate the entropy of the scalar field on the KN
black hole. Before doing it, however, there are two points to be considered and resolved. One is on the non-rotating limit of eq.(18), and the other is on the $\theta$-dependence structure of the brick wall cutoff $\epsilon$, which is especially important
 to the calculation of the entropy. We first consider the problem on 
the non-rotating limit of the free energy, because, as we shall see, the
result of it provides a rough estimation for the $\theta$-dependence structure of the brick wall cutoff $\epsilon$. 

We showed that according to the condition (15) the SR part of the free energy does not vanish only in the ranges $r_H < r < r_{erg} $, $0< x < \alpha $. Since in the non-rotating case, $r_H = r_{erg} $, the SR modes do not contribute to the free energy. T
his is a natural fact because the SR modes do not exist in the non-rotating case[15]. However, taking the limit $\Omega_H \rightarrow 0$ or $a \rightarrow 0$ into eq.(17), we can see that $F_{SR}$, precisely speaking $f_2(\theta)$ in eq.(18), becomes dive
rgent. Therefore, we encounter an apparent inconsistency. The same situation was present also in the (2+1) dimensional rotating black hole case[14].

This inconsistency is due to the assumptions that the scalar field on the KN black hole background is rotating with the horizon angular velocity $\Omega_H \neq 0$, and the inequality, $r_{erg} \geq r_H + \epsilon(\theta)$, is satisfied. The essence of the
 brick wall method is to count the modes outside the brick wall and under the assumption of the inequality the $r$-integration range in eq.(17) is well defined. But, if $\Omega_H$ is so small such that $r_{erg} \leq r_H + \epsilon(\theta)$, it happens tha
t we count the SR modes inside the brick wall. Thus in order to preserve the validity of the brick wall method for the rotating black hole, the angular velocity has to be larger than a certain value. The lower bound of the angular velocity (momentum) is d
etermined by looking at the inequality,
$r_{erg} \geq r_H+\epsilon(\theta)$, and is given by
\begin{equation}
\label{abound}
a^2 \geq \frac{2\sqrt{M^2-e^2}}{\sin^2 \theta} \epsilon(\theta) 
     -\frac{1+\cos^2 \theta}{\sin^4 \theta} \epsilon^2 (\theta)
     + {\cal O} ( \epsilon^3 (\theta)) .
\end{equation}
This fact implies that in the framework of the brick wall method, taking the non-rotating limit of the free energy is meaningless, and the non-rotating and rotating cases should be treated separately.

On the other hand, the lower bound of the angular momentum (20) tells us how the brick wall cutoff depends on $\theta$. We see that, if $\epsilon (\theta)$ does not vanish at $\theta =0, \pi$, the right hand side of the inequality (20) diverges at the poi
nts. Since, in addition to the lower bound (20), the momentum $a$ has an upper bound, $M^2-e^2 \geq a^2$, which is inherent in the metric (1), the right hand side of the inequality (20) must be regular at any $\theta$. Therefore, $\epsilon (\theta)$ has t
o vanish at $\theta = 0, \pi$. One possible solution for the regularity is to make $\epsilon (\theta)$ involve $\sin^n \theta$, $(n \geq 2)$, with some regular function depending on $\theta$ and other black hole parameters
\begin{equation}
\epsilon(\theta) = {\bar \epsilon} g(\theta)\sin^n \theta ,
\end{equation}
where ${\bar \epsilon}$ is a small constant and
$g(\theta)$ is a regular function not yet determined.

According to the $\theta$-dependence structure of $\epsilon$ (21), we can understand an important thing in the brick wall method for the KN black hole.  Firstly, in order to eliminate divergences of the scalar field states arising close to
the horizon, the `radial' cutoff $\epsilon$ is introduced. But, since $\epsilon (\theta )$ vanishes at $\theta = 0, \pi$, regardless of introducing the radial cutoff $\epsilon$, the unexpected divergence remains at the points. This story leads us to intro
ducing another cutoff 
on the $\theta$-integration range $\delta$.  Then, from the eqs.(10) and (18), the total free energy  is given by
\begin{eqnarray}
F &=& F_{NSR} + F_{SR}
\nonumber \\
&\approx& - \frac{\zeta(4)}{2\beta^{4}} \int ^{\pi -\delta}_{\delta}
d\theta 
\int^{2\pi}_{0} d\phi 
  \frac{(r_{H}^{2} + a^{2})^{4}\sin\theta }{(r_{H}^{2} + a^{2}\cos^{2}   
\theta)
 (r_{H} - M)^{2}} \frac{1}{\epsilon(\theta)} 
 + \frac{F_{1}}{\epsilon^{4}_{m}}
 + \frac{1}{\beta^{4}}{\cal O}\left( 
\frac{1}{\sqrt{\epsilon(\theta)}}
                              \right) ~.
\end{eqnarray}

\section{ENTROPY}
We are now ready to obtain the entropy of the scalar field on the KN black hole black hole background from the standard formula
\begin{eqnarray}
S = \beta^{2}  \displaystyle \left.
        \frac{\partial F}{\partial \beta} \right|_{\beta=\beta_{H}}.
\end{eqnarray}

First of all, consider the contribution of the SR modes to the entropy. As mentioned above, the total free energy $F$ is contributed by the SR modes as well as the NSR ones. The leading order contribution of the SR part is related with that of the NSR par
t as $F_{SR} \approx -F_{NSR}/2 $. Thus the total entropy is also given by $S=S_{NSR}+S_{SR}$ with
$S_{SR} \approx - S_{NSR}/2$. In other words, 
the SR modes negatively contributes to the entropy. Of course, the total entropy is positive.  This negative contribution may be understood from the fact that the modes are named as `the superradiant modes'[16] ;  It is well known that the SR scattering i
s one way that the rotational energy of a rotating black hole can be extracted, in principle.  
In the first law of 
the black hole mechanics, the viscous torque $\Omega_{H} dJ/dt$ produces 
an energy extraction in the superradiant range $0<E< m \Omega_{H}$.  This leads to the decreasing of the black hole mass.  As a result, 
the black hole entropy is diminished.  

Substituting (23) for (22), upto leading order, the entropy is given by
\begin{eqnarray}
S & \approx & \frac{2 \zeta (4)}{(2 \pi)^3} (r^2_H+a^2) (r_H-M)
         \int_{\delta}^{\pi - \delta} \int_{0}^{2\pi} d \theta d \phi 
    \frac{ \sin \theta }{ (r^2_H + a^2 \cos^2 \theta ) }
    \frac{1}{ \epsilon (\theta) } 
\nonumber \\
  &=& \frac{2 \zeta (4)}{(2 \pi)^4} A_H
         \int^{\pi -\delta}_{\delta} d \theta \int^{2\pi}_{0} d \phi 
    \frac{\sin \theta}{ {\tilde \epsilon}^2 (\theta) },
\end{eqnarray}
where we used
\begin{eqnarray}
&&A_{H} = 4 \pi (r_{H}^{2} + a^{2}),          
\nonumber \\
&&\beta_{H} = \frac{2 \pi  (r_{H}^{2} + a^{2}) }{ r_{H} - M },  
\nonumber  \\
      &&{\tilde \epsilon }(\theta) \approx   \sqrt{2} \left(
\frac{r_{H}^{2} 
  + a^{2}\cos^{2}\theta}{r_{H} -M} \right)^{1/2}\sqrt{\epsilon (\theta)}~,
\end{eqnarray} 
and $A_H$,  $\beta_H$, and ${\tilde \epsilon} (\theta)$ are the area of the black hole horizon, the inverse Hawking temperature and  the invariant distance from the horizon to the brick wall, respectively.

At present, we have no guide line for
choosing a particular value of $n$ and $g(\theta)$ in eq.(21). In this paper, we choose $n=2$ and
\begin{equation}
g(\theta) \equiv \frac{r_H-M}{2(r^2_H+a^2\cos^2\theta)}~.
\end{equation}
Then the invariant cutoff ${\tilde \epsilon}(\theta)$ (25) is given by
\begin{equation}
{\tilde \epsilon}(\theta) \approx \sin \theta \sqrt{ {\bar \epsilon} }~.
\end{equation}
In this case, the entropy (24) becomes
\begin{eqnarray}
S  \approx  \frac{4 \zeta (4)}{(2 \pi)^3} 
 \frac{1}{ {\bar \epsilon} }\ln \left(\cot \frac{\delta}{2} \right)A_H
   \approx  \frac{4 \zeta (4)}{(2 \pi)^3} 
 \frac{1}{ {\bar \epsilon} } \ln \left( \frac{\delta}{2}\right)A_H~.
\end{eqnarray}
We see in eq.(28) that the entropy becomes certainly divergent when $\delta$ as well as ${\bar \epsilon}$ goes to zero. Since $n \geq 2$ in eq.(21), one cannot take off the divergence of the entropy due to the angular cutoff $\delta$. Instead, it becomes 
more singular
as $n$ is on the increase. As a result, it is inevitable to introduce the angular cutoff $\delta$ as well as the radial cutoff ${\bar \epsilon}$.

In eq.(28), if cutoffs ${\bar \epsilon}$ and $\delta$ satisfy the relation
\begin{eqnarray}
\left(\frac{\delta}{2} \right)^{{\bar \epsilon}} 
= \exp \left(\frac{\pi}{45} \right),
\end{eqnarray}
the entropy of the scalar field becomes equal to the Bekenstein-Hawking 
entropy as follows
\begin{eqnarray}
 S = \frac{1}{4}A_{H}.
\end{eqnarray}
As expected, the relation (30) does not depend on the black hole parameters : the angular momentum, the charge, and the mass of the black hole.  Therefore, at least in our choice (26) and $n=2$, we can say that the cutoff values in brick wall model 
are intrinsic properties of 
the KN black hole.  

\section{SUMMARY}
We have studied the entropy of the KN black hole in terms of the brick 
wall method.  Especially, we have concentrated on the contribution of the 
superradiant modes, which is common feature of rotating black holes.
The free energy of the SR modes is divergent because of the large azimuthal quantum number.  This difficulty can be resolved by introducing the regulating factor $e^{-\epsilon_{m} m}$. Then the total free energy is contributed by the SR modes as well as t
he NSR ones. Fortunately, the unexpected divergence is independent of the inverse temperature $\beta $, and it does not contribute to the entropy. In the KN black hole case, the leading contribution to the entropy is given from both the SR and the NSR par
ts, while in the (2+1) dimensional rotating black hole case, the contribution of the SR modes is sub-leading order. 

The fact that there exist the contribution of the SR modes gives us some reasonable results in the context of the brick wall method. Firstly, the leading and next sub-leading terms (linear and logarithmic divergence) of $F_{SR}$ is minus a half
times those of $F_{NSR}$.  This means that the SR modes negatively contribute to the total free energy $F$ and the entropy. This fact may be understood from the procedure of the SR scattering between the scalar field and the KN black hole.  In the first l
aw of the black hole mechanics, the viscous torque $\Omega_{H} dJ/dt$ 
produces an energy extraction in the SR range $0<E< m \Omega_{H}$ due to the superradiant scattering. This leads to the decreasing of the black hole mass. Thus the black hole entropy is diminished. 

On the other hand, the free energy of the SR part (18) becomes divergent in non-rotating limit. In order to preserve the validity of the brick wall method for the rotating black hole, the angular velocity must have a lower bound.  This fact implies that i
n the framework of the brick wall method, taking the non-rotating limit of the free energy is meaningless, and
the non-rotating and rotating cases should be treated separately. 

This lower bound of the angular velocity gives the $\theta$-dependence structure of the brick wall cutoff as $\epsilon(\theta) = {\bar \epsilon} g(\theta)\sin^n \theta~(n \geq 2) $.  In this structure, $\epsilon$ goes to zero  at $\theta = 0, \pi$. Thus r
egardless of introducing the radial cutoff $\epsilon$, the unexpected divergence remains at the points. As a result, in order to  regulate the entropy at the horizon, we need to introduce the angular cutoff $\delta$ as well as the radial cutoff $\epsilon$
.

We have shown that if one chooses $n=2$ and a particular regular 
function (26), one can recognize that the entropy is proportional to the horizon area. Additionally, if the cutoff values
are satisfied a proper relation (29), the entropy becomes
equal to the Bekenstein-Hawking one.  The relation does not depend on the black hole parameters. Thus, the cutoffs can be regarded as an 
intrinsic property of the KN black hole in the brick wall context.  

\section*{Acknowledgments}
We were supported in part by Basic Science Research Institute Program, Ministry of Education, Project No. BSRI-96-2414. W.T. Kim was supported in part by the Korea Science and Engineering Foundation No. 95-0702-04-01-3
(1996). H. Shin was supported in part by the Korea Research Foundation (1996).


\begin{references}
\bibitem{[1]} J. D. Bekenstein, Phys. Rev. D{\bf 7}, 2333 (1973); {\it ibid}
           D{\bf 9}, 3292 (1974).
\bibitem{[2]} S. W. Hawking, Commun. Math. Phys. {\bf 43}, 199 (1975).
\bibitem{[3]} J. W. York, Phys. Rev. D{\bf 28}, 2929 (1983).
\bibitem{[4]} V. Frolov and I. Novikov, Phys. Rev. D{\bf 48}, 4545 (1993).
\bibitem{[5]} L. Susskind and J. Uglum, Phys. Rev. D{\bf 50}, 2700 (1994).
\bibitem{[6]} J. M. Maldacena, `Black Holes in String Theory', hep-th/9607235.
\bibitem{[7]} V. Frolov, D. Fursaev, A. Zelnikov, `{\it Off-Shell vs On-Shell}', hep-th/9512184 ; {\it ibid}, `{\it Black Hole Entropy : Thermodynamics, Statistical-Mechanics, and Subtraction Procedure}', hep-th/9603175.
\bibitem{[8]} G 't Hooft, Nucl. Phys. B{\bf 256}, 727 (1985).
\bibitem{[9]} R. B. Mann, A. Shiekh and L. Tarasov, Nucl. Phys. B{\bf 341},
           134 (1990).
\bibitem{[10]} R. B. Mann, L. Tarasov and A. Zelnikov, Class. Quant. Grav.
           {\bf 9}, 1487 (1992).
\bibitem{[11]} A. Ghosh and P. Mitra, Phys. Rev. Lett. {\bf 73}, 2521 (1994).
\bibitem{[12]} A. Ghosh and P. Mitra, Phys. Lett. B{\bf 357}, 295 (1995).
\bibitem{[13]} M. H. Lee and J. K. Kim, Phys. Rev. D{\bf 54}, 3904 (1996).
\bibitem{[14]} S. W. Kim, W. T. Kim, Y. -J. Park and H. Shin, Phys. Lett. 
	   B{\bf 392}, 311 (1997).
\bibitem{[15]} J. -G. Demers, R. Lafrance, and R. C. Myers, Phys. Rev. D{\bf 52}, 2245 (1995).
\bibitem{[16]} K. S. Thorne, W. H. Zurec and R. H. Price; R. H. Price, 
            I. H. Redmount, W. -M. Suen, K. S. Thorne, D. A. Macdonald and 
            R. J. Crowley, {\it Black Holes: The Membrane Paradigm}, edited 
	    by K. S. Thorne, R. H. Price and D. A. Macdonald 
            (Yale Univ. Press, 1986).
\bibitem{[17]} E. T. Newman, E. Couch, K. Chinnapared, A. Exton, A. Prakash, and R. Torrence, J. Math. Phys. {\bf 6}, 918 (1965).
\bibitem{[18]} J. B. Hartle and S. W. Hawking, Phys. Rev. D{\bf 13}, 2188 (1976).
\end{references}
\end{document}